\documentclass[aps,prl,prabib,twocolumn,showpacs,superscriptaddress]{revtex4}
\usepackage{psfig}
\usepackage{graphicx}
\usepackage{amsmath}
\usepackage{amssymb}
\usepackage{amsfonts}
\usepackage{color}
\usepackage{bm}

\begin{document}

\title{Matterwave localization in disordered cold atom lattices}

\author{Uri Gavish}
\affiliation{Laboratoire Kastler Brossel, \'Ecole Normale
Sup\'erieure, 24 rue Lhomond, 75231 Paris Cedex 05, France}
\affiliation{Institute for Theoretical Physics, University
of Innsbruck, 6020 Innsbruck, Austria}

\author{Yvan Castin}
\affiliation{Laboratoire Kastler Brossel, \'Ecole Normale
Sup\'erieure, 24 rue Lhomond, 75231 Paris Cedex 05, France}
\email{yvan.castin@lkb.ens.fr}

\begin{abstract}
We propose to observe Anderson localization of
ultracold atoms in the presence of a random potential
made of atoms of another species and trapped at the nodes
of an optical lattice, with a filling factor less than unity.
Such systems enable a nearly perfect experimental control of the disorder,
 while the possibility of modelling the scattering potentials by a set of point-like ones allows an exact theoretical
analysis. This is illustrated by a detailed analysis of the
one-dimensional case.
\end{abstract}



\date{23 December 2004}

\maketitle

Anderson localization \cite{electrons1} is an interference phenomenon occurring in waves
 propagating in a static disorder: rather than spreading, the wave remains
localized in a portion of space. For classical waves, localization was
observed in 2D water waves \cite{waterwave}, and in 1D \cite{Berry} and 3D \cite{Lagendijk} light beams.
When the wave corresponds to a quantum particle wavefunction, localization in a disordered
potential (of infinite spatial extension) corresponds to the
existence of square integrable stationary states at energies for which
 the classical motion is not bounded \cite{wigner}.

Indirect evidence of
Anderson localization
of electrons in condensed matter systems (e.g., the conductance dependence on the temperature) was
obtained in 2D and 3D and also in thin wires,
with an interpretation made difficult by interaction effects and the
presence of a thermal bath \cite{electrons1}. Truly 1D condensed matter systems
 are ordinarily subjected to strong interactions, and
there is presently an active debate about the role of interactions
in 2D initiated by the
 observation of a metal-insulator transition
\cite{Kravchenko} in 2D electron gases.

On the other hand, ultracold atomic gases appear as very favorable systems
for experimental study of Anderson
localization of matterwaves. These systems
have the advantage of being very flexible: due to a very weak coupling to the environment
they are virtually immune to unwanted decoherence while enabling the possibility of coupling to
a specifically engineered thermal bath \cite{Zoller} or an effective magnetic field \cite{Olshanii2}
 for the aim of probing their effect on the localization.
The Feshbach resonance
\cite{Feshbach}
allows a \emph{controllable} introduction of interactions
whose strength can be chosen at will \cite{Ketterle_feshbach},
and therefore opens the possibility of experimental tests for interaction-localization effects
as appeared in models \cite{Dorokhov} which, although analytically solvable,  were considered unrealistic until now.
Finally, the dimensionality of the gas is also adjustable by the use of tailorable optical
potentials \cite{1D2Dgas}.

A natural way of producing a disordered potential for atoms
is by using a laser speckle \cite{Salomon,Inguscio}.
This requires however a very careful
control of the speckle, to ensure that the absence of spreading
of a matterwave is not due to a trapping of an atom in a
local potential minimum, which is particularly challenging
in 3D where only the lowest energy states may be localized.
Also, comparison with
theory generally requires a numerical, rather than analytical, solution of Schr\"odinger's equation.

In this work, we propose a way to create an almost perfectly well controlled
disordered potential for an atomic matterwave, which can even be \emph{determined
by a direct measurement}.  Moreover,  this potential can be justifiably modelled by
 point-like scatterers which allows
an analytical, and often \emph{exact} \cite{Berezinksii,Smilansky} study of the localization and its observability.
Although we focus on 1D, our scheme is applicable to any dimension.
In fact, in 2D it can be subjected to an exact analysis \cite{Azbel} even in the presence
of a magnetic field.

\begin{figure}
    \begin{center}
       \includegraphics[height=3.35in,width=1in,angle=-90]{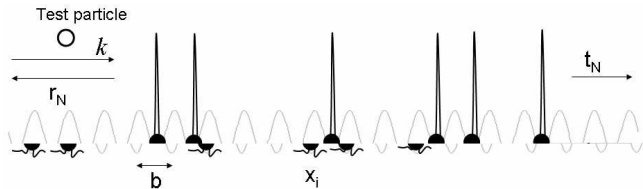}
       \vspace{-0.1cm}
         \caption{Atoms (`scatterers') trapped at the nodes of a periodic optical lattice
         and cooled to the vibrational ground state. The average occupancy is $p$ ($\approx 1/3,$ here).
         The test particle experiences only the delta potential (in black)
created by the scatterers and is blind to the periodic optical one (in grey).
} \label{fig1}
     \end{center}
\vspace{-0.85cm}
\end{figure}

We consider a gas of atoms trapped at the nodes of an optical
lattice of spatial period
$b$ (Fig.~\ref{fig1}):
each atom is cooled down to the ground vibrational state
of the local micro-well;
each node is occupied by an atom
with probability $p$ independently of the other nodes,
multiple occupancies assumed
not to occur; tunnelling
between neighboring sites is made negligible by choosing the
modulation depth of the optical lattice much larger than the energy
$\hbar^2(\pi/b)^2/m_s,$ where $m_s$ is the trapped atom mass,
to ensure that
the spatial configuration is static \cite{Bloch}.

The set of trapped atoms, designated below as `scatterers', will act as a random potential for
atoms of another species or of the same species but in
another internal state, denoted as `test particles'.
One should ensure that the test particles (unlike the scatterers) will not be trapped by the optical lattice.
This can be achieved by using two different species
with sufficiently different resonance
frequencies \cite{even_better}.
In what follows, we assume that the test particles
experience as an external potential \emph{only} the interaction potential
with the scatterers. To ensure elastic scattering
we also require that the incoming kinetic energy of a
test particle is less than the level spacing  $\hbar\omega_s$
of a trapped scatterer,
\begin{equation}
\label{eq:elastic}
\frac{\hbar^2 k^2}{2m_t} \ll \hbar \omega_s
\end{equation}
where $k$ and $m_t$ are the wavevector and the mass
of the test particle. $\omega_s$ is the oscillation frequency
of a scatterer in the local micro-well of the lattice.
Eq.(\ref{eq:elastic}) ensures, by energy conservation, that at the end of a scattering event the
 scatterer is not left in an excited vibrational level:
hence, the disordered potential is static.

Focusing on 1D we assume that the test particle is strongly
trapped in a matterwave
guide, with a quantum of vibrational energy
$\hbar \omega_t$ much larger than the longitudinal
kinetic energy,  so that its transverse $y$-$z$ motion is frozen
in the ground vibrational state of the guide.
We then introduce the model Hamiltonian for the quantum motion of a
test particle along the lattice direction $x$:
\begin{equation}
H = \frac{p_x^2}{2m_t} + \sum_{j=1}^{N} g\, \delta(x-x_j).
\end{equation}
Here the $x_j$ are the positions of the occupied
micro-wells, all integer multiple of the lattice spacing $b$.
The effect of each scatterer is represented by a Dirac delta potential
with a coupling constant $g$. This assumption is reasonable
when the wavevector $k$ of the test particle is small enough:
assuming for simplicity that $m_s=m_t=m$,
$1/k$ should be larger than the sizes $a^{\rm ho}_{s,t}
\equiv (\hbar/m\omega_{s,t})^{1/2}$
of the harmonic oscillator ground state
of a scatterer in a micro-well
and of the transverse guide ground state of the test particle \cite{Olshanii},
conditions already ensured
by the elasticity condition Eq.(\ref{eq:elastic}) and by the 1D nature
of the motion of the test particle.
A scatterer may then be modelled by a zero range potential.
When the 3D scattering length, $a,$ describing the free space
interaction between a scatterer and a test particle is much smaller
than the harmonic oscillator lengths $a^{\rm ho}_{s,t}$ one is
in the so-called Born regime and the
1D coupling constant $g$ is given by:
\begin{equation}
g= 4\hbar \frac{\omega_t\omega_s}{\omega_s+\omega_t} a.
\end{equation}

Our model Hamiltonian was shown in \cite{Ishii} to lead to
localization as a consequence of a theorem derived in Ref. \cite{Furstenberg}.
An acceptable quantitative measure of localization, that we adopt in this paper, is the decay length
of the transmission coefficient: taking
 all the scatterers to be in the half space $x\ge 0$ and introducing
the transmission amplitude $t_N(k)$ of an incoming plane wave of momentum
$k>0$ through a set of $N$ scatterers,
we define the localization constant $\kappa(k)$ as
\cite{Ziman}:
\begin{equation}
\kappa(k) = \lim_{N\rightarrow +\infty} \langle \frac{-\log |t_N(k)|}{x_N-x_1}
\label{eq:kappa}
\rangle.
\end{equation}
The average $\langle .. \rangle $ is over all possible realizations of the disorder;
although, strictly speaking it is  \emph{not required} since
 $\log |t_N(k)|$, contrarily to $|t_N(k)|$ itself,
is a self-averaging quantity for
$N\rightarrow +\infty$ \cite{Anderson2,Erdosh}. The transmission and reflection amplitudes are given by
\begin{equation}\label{eq:tnrn}
t_N =1/(R_N)_{11}^* \ \ \ r_N/t_N=-(R_N)_{21},
\end{equation}
where $R_N$ is related to the transfer matrix of the matterwave through $N$ scatterers, $\tilde{R}_N,$ through the relation
\begin{eqnarray}\label{eq:RN}
R_N(x_1,\ldots,x_N) \equiv T(x_N)^{-1} \tilde{R}_N T(x_1)~~~~~~~~~~~~\\ \nonumber
=G_0 T(x_N-x_{N-1})^{-1} G_0 \ldots T(x_2-x_1)^{-1} G_0.
\end{eqnarray}
$G_0$ is the transfer matrix of a single scatterer at $x=0$:
\begin{equation}\label{eq:G0}
G_0 = \left(
\begin{matrix}
1-i\alpha & -i \alpha \\
i\alpha & 1+i\alpha
\end{matrix}
\right)
\end{equation}
with $\alpha = m g/(\hbar^2 k),$ and $T(x)$ is the transfer
matrix corresponding to a free propagation over an
abscissa $x$:
\begin{equation}
T(x) = \left(
\begin{matrix}
e^{-ikx} & 0 \\ 0 & e^{ikx}
\end{matrix}
\right).
\end{equation}
The use of $R_N$ instead of $\tilde{R}_N$ in Eq. (\ref{eq:tnrn}) does not affect $|t_N(k)|,$
but simplifies the calculations,
 since $R_N$ depends only on the variables $x_{i+1}
-x_i$, which are independent random variables with a common probability
distribution given by: $P_(s_i) = p (1-p)^{s_i-1}$ where $s_i=(x_{i+1}
-x_i)/b.$

We calculated the localization constant $\kappa$
numerically, by a Monte Carlo averaging over the disorder, taking
a large enough number of scatterers to ensure convergence in
Eq.(\ref{eq:kappa}).
When expressed in units of $1/b$, $\kappa$
 depends on three dimensionless parameters:
the filling factor $p$, the reduced momentum $kb$ and the reduced
coupling constant $m g b/\hbar^2$. Assuming for simplicity that
$\omega_s=\omega_t=\omega$, so that the harmonic oscillator
lengths also coincide, $a^{\rm ho}_s=a^{\rm ho}_t=a^{\rm ho}$,
and introducing the recoil energy $E_R=\hbar^2(\pi/b)^2/2m$, we
find $mgb/\hbar^2= 2\pi (a/a^{\rm ho}) (\hbar\omega/2 E_R)^{1/2}$.
Typically,  $\hbar\omega< 10 E_R.$ Since we
required $a/a^{\rm ho}\ll 1$, one should therefore have $m g b/\hbar^2 \ll 15$. In
all our numerical calculations $m g b/\hbar^2 =
2.278$.

\begin{figure}
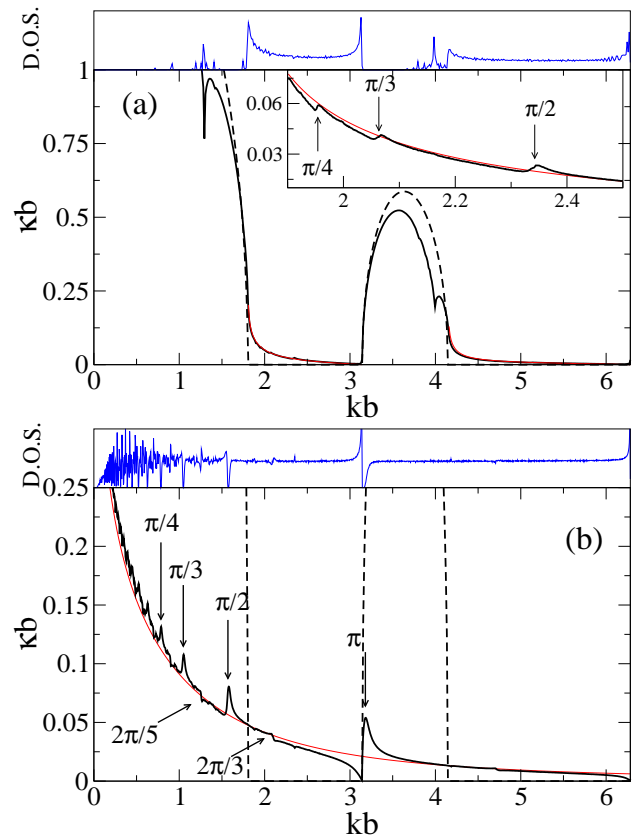

\begin{center}
\includegraphics[height=5.5cm,width=8.25cm,clip]{fig2a.eps} \\
\includegraphics[height=5.5cm,width=8.25cm,clip]{fig2b.eps}
\end{center}
\vspace{-0.5cm}
\caption{The localization constant $\kappa$ as a function of the
incoming wavevector $k$, for average occupancies
(a) $p=0.9,$  (b) $p=0.1$. The coupling
constant is $mgb/\hbar^2=2.278$. Black solid lines: numerics.
Red thin line:   analytical result for
$1-p\ll 1$ (a), and $p\ll 1$ (b).
Peaks and steps in $\kappa$ marked by arrows
in the inset of (a) and
in (b), with the corresponding values of $\bar\theta$ for (a)
and of $kb$ for (b) being rational multiples of $\pi$.
Blue line in upper frames of (a),(b): density of states.
Dashed lines in (a): $\kappa$ for the periodic case $p=1$.
} \label{figk}
\vspace{-0.5cm}
\end{figure}

 Figure \ref{figk} shows the localization constant dependence
on the momentum $k:$ Fig. \ref{figk}a for filling factor $p=0.9$ (solid lines) and $p=1$ (dashed line),
and Fig. \ref{figk}b for filling factor $p=0.1$ (solid lines).
In the case $p=1$ where the scatterers form a finite periodic chain,
allowed bands (where $\kappa = 0$
corresponding to infinite localization length of the Bloch waves)
are separated by forbidden gaps.
The $p=0.9$ case corresponds to a perturbation
of the periodic chain by an occasional appearance of empty sites.
Now $\kappa$ takes non-zero values
in the former allowed bands while its value in the former forbidden gaps
is only weakly affected by the disorder.
For $p=0.1$ the band structure of the periodic chain is
washed out, except for the vanishing of $\kappa$ in the points
where $kb$ is an integer multiple of $\pi$ (corresponding to
band edges in the periodic case).
 The peaks and steps on the $\kappa$ curves (marked
by arrows) will be interpreted
analytically below, in terms of a phase shift being
a rational multiple of $\pi$.

At a given scattering energy, measuring
a  finite value of $\kappa$  is, strictly speaking,
not a proof of localization  but may be due simply to the presence
of a spectral gap \cite{Ziman} as is the case in periodic systems.
To confirm the existence of localized states one should calculate the density
of states and check that it is also finite.
We performed such a numerical check,
 (shown in the upper frames of Figs. \ref{figk}a and \ref{figk}b),
 by imposing periodic boundary conditions  in a box of
size $x_N-x_1,$ and averaging over disorder.

We also performed an analytic calculation of $\kappa$ in several limiting
cases. The first limiting case is the $p\rightarrow 0$ limit for
a fixed value of $kb$ and can be treated along the lines of Ref. \cite{Berry}.
By expanding the matrix product involving the two factors
$G_0 T(x_N-x_{N-1})^{-1}$ and $R_{N-1}$ in
Eq.(\ref{eq:RN}), one obtains the recursion relation:
\begin{eqnarray}
\log |t_N| & = & \log |t_0 t_{N-1}| \nonumber \\
&-&\log \left|1+r_0^* r_{N-1}\frac{t_{N-1}^*}{t_{N-1}}e^{-2ik(x_N-x_{N-1})}
\right|
\end{eqnarray}
where $r_0$ and $t_0$ are the reflection and transmission amplitudes
for the transfer matrix $G_0$. When $p$ tends to zero,
the accumulated phase shift $kb s$ (with $P(s) = p (1-p)^{s-1}$ as above) between two successive scatterers
is uniformly distributed between $0$ and $2\pi$ (modulo $2\pi$), as long as $kb$ is not a rational multiple of
$\pi$. Noting that $\int_0^{2\pi}d\theta \log|1-ze^{i\theta}|=0$ for $|z|<1,$ we obtain
\begin{equation}
\langle  \log |t_N| \rangle =
\log|t_0| + \langle  \log |t_{N-1}| \rangle
\label{eq:lowp}
\end{equation}
which leads to
\begin{equation}
\kappa b = (1/2)p\log(1+\alpha^2).
\label{eq:lowpfinal}
\end{equation}
Fig.~\ref{figk}b shows a good agreement of Eq. \ref{eq:lowpfinal} with numerics, except for the peaks at
$kb=\pi/(s+1)$ and the steps at $kb=2\pi/(2s+1)$, for $s$ integer.
The calculation is readily extended to the $p\rightarrow 1$
limit, by considering the defects in the chain as scatterers
on top of a periodic background: the propagation of the matter waves
in between two consecutive defects of distance $s b$ is given
by the transfer matrix $\bar{T}(sb)=(T(b)^{-1} G_0)^s,$ and the
scattering over a defect corresponds to the transfer matrix $\bar{G}_0=
T(b)^{-1}$. Assuming that the incoming energy is in an
allowed band of the periodic lattice, the matrix $T(b)^{-1} G_0$
has eigenvalues of the form $e^{\pm i\bar{\theta}}$. When $\bar{\theta}$ is
not a rational multiple of $\pi$, the procedure  of the previous paragraph
can be reused, replacing $G_0$ by $\bar{G}_0$ and $T(sb)$ by
$\bar{T}(sb)$. We then get a formula similar to Eq.(\ref{eq:lowp}),
with $t_0$ replaced by $\bar{t}_0$, which is the transmission
amplitude for
the transfer matrix $\bar{G}_0$ in the basis where $\bar{T}(b)$
is diagonal. This leads to
\begin{equation}
\kappa b \sim (1-p) \log \left|\frac{\exp(ikb)-\rho\exp(-ikb)}{1-\rho}\right|
\label{eq:p=1}
\end{equation}
where $\rho=|\exp(i\bar{\theta})-(1-i\alpha)\exp(ikb)|^2/\alpha^2$
and where $\bar{\theta}$ is the solution of $\cos\bar{\theta}=\cos(kb)+\alpha\sin(kb)$
for $\rho<1.$
This expression agrees well with numerics, see Fig. \ref{figk}a,
except for the peaks:
as shown in this figure, these indeed correspond to values of
$\bar{\theta}/\pi$ that are rationals.

The third limit we investigated analytically is that of
a narrow distribution of the phase shift $\theta =
s k b$ between two consecutive scatterers: denoting its average
by $\langle \theta\rangle,$ it is assumed that the
probability of finding $\theta-\langle \theta\rangle$ out of an interval
of size $\ll 1$ is small so that $kb < 1$ and
that the variance $\Delta \theta^2 = (1-p) (kb/p)^2 < 1.$
Along the lines of \cite{Baluni}
we expand the relation (derived in \cite{Ishii}):
\begin{equation}\label{furst1}
\kappa b = p\int_0^{2\pi} d\phi\, \mu(\phi)\,
\langle \log|M_{11}(\theta)e^{i\phi}+M_{12}(\theta)e^{-i\phi}|\rangle\,
\end{equation}
between $\kappa$ and the invariant
(Haar) measure $\mu(\phi)$ which is a solution to the Dyson-Schmidt equation:
\begin{equation}\label{furst2}
\mu(\phi) = \langle \mu[\gamma(\phi,\theta)]
\partial_{\theta}\gamma(\phi,\theta)\rangle
\end{equation}
where $M(\theta)=G_0 T(\theta k^{-1})^{-1}.$
$\gamma(\phi,\theta)$ is the argument of the complex number
$M_{11}(\theta)e^{i\phi}+M_{12}(\theta)e^{-i\phi}$.
It is useful to note \cite{Baluni} that Eqs. (\ref{furst1},\ref{furst2}) are invariant under
any $\theta$ independent $SU(1,1)$ similarity transformation $M(\theta)
\rightarrow D^{-1} M(\theta) D$. Assuming
the energy is in an allowed band of a periodic chain
of scatterers of period $\langle \theta\rangle/k=b/p,$
the transfer matrix $M(\langle \theta\rangle)$ has then
unimodular eigenvalues and we choose $D$ so that
$D^{-1} M(\langle\theta\rangle) D$ is diagonal. We then expand
all functions of $\theta$ in powers of $(\theta-\langle\theta\rangle)$
up to fourth order,
which requires an expansion of
$\mu(\phi)$ and $\kappa$ to first and second orders in $\Delta \theta^2$:
\begin{equation}
\mu(\phi) = \mu^{(0)}(\phi)+\mu^{(1)}(\phi)+\ldots\, ; \ \
\kappa= \kappa^{(0)} + \kappa^{(1)} + \kappa^{(2)} + \ldots
\end{equation}
One has $\mu^{(0)}(\phi)=1/(2\pi)$ and $\kappa^{(0)}=0$.
Denoting $a_r\equiv Re(e^{ikb/p}(1-i\alpha)),$ $a_i\equiv Im(e^{ikb/p}(1-i\alpha)),$ we get  \cite{tech_uri}:
\begin{eqnarray}
\label{eq:kappa1}
\kappa^{(1)}b &=& \frac{p(1-p)\alpha^2/2}{1-a_r^2}\left(\frac{kb}{p}\right)^2 \\
\kappa^{(2)}b&=& -p(1-p)^2\alpha^2\frac{a_i^2+\alpha^2/2}{1-a_r^2}\left(\frac{kb}{p}\right)^4
\\
&-& \frac{p(1-p)}{12}(p^2-9p+9)\alpha^2\frac{3a_i^2+a_r^2-1}{(1-a_r^2)^2}\left(\frac{kb}{p}\right)^4
\nonumber
\end{eqnarray}

In conclusion, we proposed a well controlled way of producing a disordered potential
for atomic matter waves, by the scattering of test particles
on scatterers trapped at the nodes of an optical lattice.
We showed how
a transmission experiment through such a one-dimensional
disordered chain  provides a clear direct evidence of Anderson localization
at energies where the density of states is appreciable.
The proposed experiment
is one of several possibilities, such as the measurement of the (absence of)
spreading of a wavepacket initially prepared inside the disordered medium
which is linked to other exactly calculable \cite{Berezinksii,Smilansky} aspects of localization:
the finite return probability and
finite inverse participation ratio. The proposed scheme is extendable
to higher dimensions
and enables a controlled experimental study of how the localization is affected by
the introduction of
 engineered thermal-like bath \cite{Zoller}, effective magnetic field \cite{Olshanii2}, interactions among test particles \cite{Dorokhov},
or several coupled channels for the transverse motion of the
test particle \cite{coupled_channels}.

We akcnowledge I. Carusotto, P. Massignan and Y. Imry
for useful discussions. Laboratoire Kastler Brossel is a Unit\'e de
Recherche de l'\'Ecole Normale Sup\'erieure et de l'Universit\'e Paris
6, associ\'ee au CNRS.
U. G. acknowledges financial support from the Cold Quantum Gases European
Network contract no. HPRN-CT-2000-00125.


\begin{thebibliography}{99}
\bibitem{even_better}
If the scatterers and the test particles are
two different internal states of a same species, one may use
a combination of far detuned laser beams with opposite detunings,
one with a circular polarization and the other with a linear polarization,
with intensities in a ratio such that the scalar light shift
produced by the linearly polarized beam compensates the light shift
of the circularly polarized beam for one of the two considered
internal states.
For alkali atoms with a wide enough fine structure in the excited
state, one can even use a single laser beam detuned in between the two
excited fine states at a `magic' detuning such that the lightshifts
compensate for one of the internal ground state sublevels.
\bibitem{tech_uri}
Taking $p\rightarrow 1$ and $kb\rightarrow 0$, with $\alpha/(kb)$
fixed and $<1/2$, we confirmed (up to order $(kb)^4$) the consistency of Eq.(\ref{eq:p=1})
and Eqs.(\ref{eq:kappa1},17). Close to the band edges $a_r\rightarrow 1,$ $\kappa^{(2)}>\kappa^{(1)}$
and the expansion is no longer justified.
\end{thebibliography}
\end{document}